\def\preprint{1}                
\def\comment#1{}

\if\preprint1
        \documentclass[usenatbib]{mn2e}
        \usepackage{natbib}
        \usepackage{times}
        \usepackage{graphicx}
        \usepackage{calc}
        \usepackage{rotating}
         \usepackage{lscape}
         \bibpunct{(}{)}{;}{a}{}{,} 
         \usepackage{longtable}
         \usepackage{tgcursor}
         \usepackage{enumitem} 
      
\else
        \documentstyle[astrop-bib,referee,times]{mn2e}
        \newcommand{\includegraphics}[1]{}
\fi

\def\oversim#1#2{\lower0.5pt\vbox{\baselineskip0pt \lineskip-0.5pt
     \ialign{$\mathsurround0pt #1\hfil##\hfil$\crcr#2\crcr\sim\crcr}}}

\title[Magnetic fields in Post-AGB stars]{First detection of surface magnetic fields in Post-AGB stars : the cases
of U Monocerotis and R Scuti}
\author[L. Sabin et al.]{L. Sabin$^{1,2}$\thanks{E-mail:laurence.sabin@gmail.com(LS)}, G.A. Wade$^{3}$ and A. L\`ebre$^{4}$ 
\\
$^{1}$ Instituto de Astronom{\'i}a y Meteorolog{\'i}a, Departamento de F{\'i}sica, CUCEI, Universidad de Guadalajara, Av. Vallarta 2602, C.P. 44130, Guadalajara, Jal., Mexico\\
$^{2}$ Instituto de Estudios Avanzados de Baja California, A. C., Av. Obreg\'on 1755, 22800 Ensenada, BC, Mexico.\\
$^{3}$ Department of Physics, Royal Military College of Canada, PO Box 17000, Station Forces, Kingston, Ontario K7K 7B4, Canada.\\
$^{4}$ Laboratoire Univers et Particules de Montpellier - CNRS and Universit\'e Montpellier II - Place E. Bataillon, 34 090 Montpellier, France\\
}
\begin{document}

\date{Accepted 2014 October 10.  Received 2014 October 7; in original form 2014 September 11}

\pagerange{\pageref{firstpage}--\pageref{lastpage}} \pubyear{2014}

\maketitle

\label{firstpage}

\begin{abstract}

While several observational investigations have revealed the presence of magnetic fields in the circumstellar envelopes, jets and outflows of post-Asymptotic Giant Branch stars (PAGBs) and planetary nebulae (PNe), none has clearly demonstrated their presence at the stellar surface. The lack of information on the strength of the surface magnetic fields prevents us from performing any thorough assessment of their dynamic capability (i.e. material mixing, envelope shaping, etc). We present new high resolution spectropolarimetric (Stokes $V$) observations of a sample of PAGB stars, realised with the instruments ESPaDOnS and Narval, where we searched for the presence of photospheric magnetic fields. Out of the seven targets investigated the RV Tauri stars U Mon and R Sct display a clear Zeeman signature and return a definite detection after performing a least squares deconvolution (LSD) analysis. The remaining five PAGBs show no significant detection. We derived longitudinal magnetic fields of $10.2\pm 1.7$~G for U Mon and $0.6\pm 0.6$~G for R Sct. In both cases the Stokes profiles point towards an interaction of the magnetic field with the atmosphere dynamics. This \textit{first discovery of weak magnetic fields} (i.e. $\sim$10 gauss level) at the stellar surface of PAGB stars opens the door to a better understanding of magnetism in evolved stars.

\end{abstract}


\section{Introduction}

The magnetic field strength at the surface or photosphere of evolved intermediate mass stars (with initial masses in the range $\sim$~0.8--8 M$\odot$) such as Asymptotic Giant Branch (AGB) stars, Post Asymptotic Giant Branch (PAGB) stars and Planetary Nebulae (PNe) is a poorly if not totally unknown parameter to this date. This knowledge is nonetheless of great importance for the understanding of the role played by magnetic fields in the dynamics as well as in the chemistry of these stellar objects.\\
In the case of AGB stars, the capacity of the field (or the Alfv\'en waves) to constrain and/or govern the mass loss is an important aspect regarding the subsequent evolution (see below). Magnetic fields have been detected and measured in the envelopes of AGB stars via OH, H$_{2}$O and SiO maser studies \citep{Rudnitski2010,Vlemmings2005,Herpin2006} and linear polarisation of molecular lines such as CO, SiS and CS \citep{Vlemmings2012,Girart2012}. Magnetic fields have also been detected at the surface of AGB stars via circular spectropolarimetry (\citealt{Konstantinova2013,Konstantinova2014}; \citealt{Lebre2014}). 
The latter authors obtained the first detection of a magnetic field at the stellar surface of a (S-type) Mira star, namely $\chi$ Cygni, and derived a weak longitudinal magnetic field strength of 2--3 gauss. The association of the Zeeman signature with only one of the pulsationally doubled line components suggests a relation between magnetism and atmospheric dynamics, in particular shock waves. \citet{Konstantinova2013} have detected weak magnetic fields (from 1 to 10 G) at the surfaces of bright single M-type giants located at the tip of the Red Giant Branch (RGB) and along the AGB (semi regular stars). For some cases, an intermittent magnetic field is reported indicating a non uniform surface distribution of the magnetic regions or an intrinsically-variable field.\\
Regarding the subsequent evolutionary phases, it is well established now that the majority of PAGB stars and PNe display non-spherical geometries \citep{Manchado1997,Sahai2011,Parker2014} and although the action of a (close) stellar companion is often invoked as the principal explanation of the observed morphologies \citep{DeMarco2013,Jones2011,Miszalski2011}, magnetic fields should not be discarded as various studies have unveiled their presence in the circumstellar shells or nebulae. Thus, there are now several evidences suggesting an active role of the magnetic field in the circumstellar environment. We can cite the detection of large scale magnetic fields in the environment of PAGB stars and PNe via dust polarisation analysis by \citet{Greaves2002} and \citet{Sabin2007,Sabin2014}, the detection of synchrotron emission in the PAGB star IRAS 15445-5449 by \citet{Perez2013}, the detection of a magnetically collimated jet in the evolved star W43A by \citet{Vle2006}, the measurement of linear and/or circular polarisation of maser emission by e.g. \citet{Etoka2009}, \citet{Ferreira2013} and \citet{Amiri2010} and the presence of filamentary structures \citep{Huggins2005}. 
 In particular, several of the aforementioned investigations underlined the ability of the field to launch and collimate outflows. In this case weak fields (of a few gauss) are sufficient for these processes to occur \citep{Tocknell2014}. \\
 If in several cases it was possible to unveil the (large scale) geometry or distribution of the magnetic field (e.g. via dust polarisation analysis), this is not generally the case for the estimation of its global strength or intensity at the stellar surface or in the envelope. 
 A method to derive the magnetic intensity at the stellar surface consists in measuring the magnetic intensity from circumstellar masers located at known distances (r) from the central star. Depending on which field configuration\footnote{B $\propto$ r$^{-1}$, B $\propto$ r$^{-2}$, B $\propto$ r$^{-3}$ for a toroidal, poloidal or dipole structure respectively which are the configurations generally assumed (see the models by \citet{Blackman2001} and \citet{Pascoli2010} for example).} is best fitted by the masers distribution in a Log $B$ {\it vs} Log $r$ graph, one could extrapolate the field's strength at the stellar surface. For instance, \citet{Ferreira2013} used this technique to infer the magnetic strength at the surfaces of a sample of AGB stars and deduced fields ranging from the weak regime i.e. lower than $\simeq$ 10 G (mostly assuming a toroidal field configuration, to an intermediate regime from $\sim$2 to $\sim$115 gauss (assuming a poloidal field) and up to the kilogauss level (assuming a dipole field). The recent spectropolarimetric studies dedicated to AGB stars by \citet{Konstantinova2013} seem to favour weak (global) surface fields.\\
\indent A more unequivocal and effective approach is the direct measurement of the field in the central star via the study of the longitudinal Zeeman effect as expressed in circularly polarized light and measured through spectropolarimetric observations. Historically however, for objects located beyond the AGB, different investigations conducted with this technique, principally targeting PNe, led to null or inconclusive detections \citep{Leone2011,Jordan2012,Leone2014,Steffen2014}. The limited number of suitable spectral lines are often to blame when analysing PNe.\\
\indent To overcome the lack of (observational) information, we present in this paper a new spectropolarimetric study of a sample of evolved intermediate mass stars, all being PAGB stars. In $\S2$ we specify the characteristics of the selected sample, in $\S3$ we present the observations and data reduction process; the result of the field measurements are presented in $\S4$, and finally the discussion and concluding remarks are presented in $\S5$ and $\S6$ respectively.\\

\begin{table*}
\caption[]{\label{Sample} Basic characteristics and stellar parameters of the PAGB sample }
\hspace{-1cm}
\begin{tabular}{|c|l|c|l|l|c|l|c|c|c|}
\hline
 HD  & Other name &   V mag & Class & Spectral &  Binary & T{\tiny{eff}}  & log g  & $\left[Fe/H\right]$  & References\\ 
   &  &    &  &  type$^{\dagger}$&   &  (K) &  (cgs) & (dex)   &\\ 
\hline
  56126  & CY CMi &   8.3 & IRexc& F5Iab     &   --       & 6500   & 0.5   & $<$-1.0   &  a,b                 \\ 
  59693  & U Mon  &  5.8 & RV Tau & K0Ibpvar & yes        & 5000   & 0.0   & -0.8      &  c,d               \\
  161796 & V814 Her & 7.2& UU Her &  F3Ib    & --         & 6600   & 0.25  & -0.3      &  e,f                  \\  
  163506 & 89 Her &  5.4 & UU Her & F2Ibe    & yes        & 6500   & 1.0   & 0.0/-0.4  &  f,g,h              \\ 
  170756 & AC Her &  7.0 & RV Tau & F4Ibpvar & yes        & 5500   & 0.5   & -1.5      &  i           \\ 
  172481 & V4728 Sgr & 9.1 & IRexc & F2Ia0   & yes        & 7250   & 1.5   & -0.55     &  j            \\
  173819 & R Sct &  5.2 & RV Tau & K0Ibpv    & --         & 4500   & 0.0   & -0.4      & d            \\
\hline
\end{tabular}
\begin{minipage}{13.5cm}
$^{\dagger}$ From \citet{Szczerba2007}. (a) \citet{Parthasarathy1992}, (b) \citet{Barthes2000}, (c) \citet{Pollard1995}, (d) \citet{Giridhar2000}, (e) \citet{Stasinska2006} , (f) \citet{Luck1990}, (g) \citep{Ferro1984}, (h) \citet{Waters1993}, (i) \citet{Winckel1998}, (j) \citet{Reyniers2001}. 
\end{minipage}
\end{table*}

\section{The sample}

\begin{table*}
\begin{center}
\caption[]{\label{Obs} Observing log of the sample presenting for each object the observing date, number of Stokes $V$ sequences, length of the sub-exposures and peak signal-to-noise ratio. }
\hspace{-0.5cm}
\begin{tabular}{|l|c|c|c|c|c|c|c|}
\hline
 Parameters &  CY CMi   & U Mon & V814 Her & 89 Her & AC Her & V4728 Sgr  & R Sct\\
 \hline 
 Obs date & 2014-02-11 & 2014-04-10 & 2009-02-16 & 2014-06-15/19 & 2014-06-19 & 2014-06-16  & 2014-07-21/23\\
 N$^{\circ}$ Stokes $V$ seq. & 2 & 5 & 2 & 8 & 4 & 2  & 6 \\
 Subexp. time (s) & 800 & 284 & 600 & 245 & 800 & 800  & 300\\
 Peak S/N & 589 &1073 &1028 &958 &900& 885 &  1000-1500 \\
\hline
\end{tabular}
\end{center}
\end{table*}

\noindent With no previous extensive spectropolarimetric studies dedicated to PAGB stars, it is difficult to estimate a priori the likelihood for these evolved objects to host surface magnetic fields and even more to quantitatively predict the characteristics of such  fields. With other research groups already dedicated to the study of PNe (see above), we therefore decided to focus our first investigation on the photospheres of bright PAGB objects and tackle the detection problem upstream, using high resolution spectropolarimeters. \\
\indent PAGB stars are at the stage of stellar evolution when the large mass loss events have mostly stopped and the temperature of the central star is not sufficiently high to fully ionise the surrounding envelope (see review by \citealt{Winckel2003}). Therefore for these objects it is possible to obtain spectra of the central stars that are more weakly contaminated by (mostly) nebular emission lines. This is a great advantage, since in the case of PNe, the numerous characteristic emission lines of the observed spectrum seriously hamper a proper magnetic analysis (see \citealt{Leone2014}). With the same purpose of avoiding contamination, we chose objects not embedded in (dense) circumstellar material. Another selection criterion was the objects' brightness, which we limited to $\simeq$10 mag (in V band) in order to obtain a sufficient signal to noise ratio (S/N) for each frame taken. We therefore set as a lower limit an estimated peak S/N $\simeq$ 500-1000 in the red part of the stellar spectrum (also to avoid local CCD saturation in the reddest orders). \\
We focused mainly on late spectral type PAGB stars (between F5 and K0) as they will ensure the presence of well defined, sharp absorption profiles as well as rich metallic-line spectra (while avoiding severe blending by molecular absorption, as it is the case in spectra of cooler AGB stars). These features will allow us to perform  an accurate spectropolarimetric analysis extracting information from several thousands of atomic lines (see \S4).\\
Finally, in order to investigate the possible connection between the occurrence of magnetic field and parameters such as the stellar class and the presence of a binary companion, we divided our sample first into different classes: ``simple" Infra Red excess stars (IRexc), UU Herculis stars (UU Her) and RV Tauri stars (RV Tau). Then we subdivided this sample into binary and single stars. All these characteristics could be important to determine the factors linked to magnetic activity and would therefore allow us to make predictions on the presence of magnetic fields. \\
\indent {\it i)} Due to the amount of material (principally dust) expelled during the AGB phase and the dust thermal emission, PAGB stars are expected to show an infra red excess. The objects classified as ``simple Infra Red excess'' stars do not show signs of large amplitude variations in their light curves but often display irregularities. 
RV Tauri stars are highly luminous variable giants/supergiants in the PAGB phase \citep{Winckel1999}. The changes in luminosity as well as spectral type (from F to K) are usually linked to stellar pulsations and occur on periods of 30--150 days. These cycles are however not always regular and can change from one to the next. The optical light curves, which show alternating deep and shallow minima, show magnitude changes up to $\Delta$V$\sim$4 mag. RV Tauri stars are also known to exhibit large IR excess and are believed to host a circumstellar disc as well as a companion star within a binary system \citep{Ruyter2005}. Finally, UU Herculis-type stars have been described by \citet{Sasselov1983,Sasselov1984} as supergiants characterised by semiregular variations with two distinct periods of 30--90 days. Thus, the pulsations of these stars appear to stop suddenly before re-starting after a couple of months. The light curves of UU Her stars show small amplitude variations of the order of $\sim$0.3 mag in the V band.\\ 
\indent{\it ii)} Among the factors associated with the presence and ``shaping action'' of magnetic fields in evolved stars is  binarity. \citet{Nordhaus2007} have estimated the need for a stellar companion within a common envelope to sustain magnetic fields until the PN phase. The dynamo created via differential rotation is thought to generate a field sufficiently efficient to shape the PAGB and PNe through magnetically collimated ejection of their wind/envelope for example \citep{Tocknell2014}. We point out that rotation in single stars has been discarded by \citet{Garcia2014} as a mechanism leading to bipolar PNe. While convection remains a possible shaping agent (via the production of a dynamo), the difference between the high number of AGB stars with convective envelopes and the lower number of bipolar PNe indicates the need for an additional parameter.\\

To balance the considerations above, and to define a sample of sufficient size and breadth to provide useful information about the general properties of magnetic fields in PAGB stars, we selected seven well-known objects, namely CY CMi, U Mon, V814 Her, 89 Her, AC Her, V4728 Sgr and R Sct\footnote{We emphasize that although we consider R Sct as a genuine member of the PAGB group, this star is also suspected by some authors to be a thermal-pulsing AGB star \citep{Matsuura2002}}. The objects and some of their characteristics are presented in Table \ref{Sample} and were all selected from the ``Toru\'n catalogue of Galactic PAGB and related objects'' \citep{Szczerba2007,Szczerba2012}. We also indicate in the same table the stellar parameters of interest for the data analysis.\\

\begin{itemize}[noitemsep,topsep=0pt]
\item CY CMi (= HD 56126) has been studied by \citet{Oudmaijer1994} and \citet{Klochkova2007} and followed in detail by \citet{Barthes2000} and \citet{Lebre2001}. All the authors reported strong changes in the pulsation of the star. The latter derived a pulsation period of 36.8$\pm$0.2 days and described the pulsation as a ``chaotic signal''. The photometric variability is particularly low, with  amplitude change between 0.06 mag and 0.15 mag (in the V band). With this information, it is difficult to establish any ``best'' observing date and CY CMi was the only object from our sample for which observations were programmed with no time constraint.
\item  U Mon (= HD 59693), which is a well known RV Tauri star, exhibits a pulsation period of 92.26 days and an amplitude variability of 2.2 mag in the V band (see the American Association of Variable Star Observers catalogue, AAVSO\footnote{http://www.aavso.org/}).
\item The amplitude variation observed for the UU Herculis star V814 Her (= HD 161796) is 0.15 mag (AAVSO) and periods of 43 and 62 days were identified by \citet{Fernie1983} and \citet{Percy1981} respectively.
\item The UU Herculis PAGB 89 Her (= HD 163506) displays the complexity typical of all the members of its class. Indeed, if a period of 68 days has been reported by \citet{Fernie1981} and later set to 65.2 days by \citet{Percy2000} (with an additional longer period of 283 days), the light variations can disappear abruptly and be replaced by erratic random fluctuations of $\Delta$V$\sim$0.05 mag.
\item AC Her (= HD 170756) is a RV Tauri star with a period of 75.29 days and a significant magnitude change $\Delta$V=2.31 mag (AAVSO and General Catalogue of Variable Stars: GCVS\footnote{http://www.sai.msu.su/gcvs/gcvs/}). The pulsations also lead to a variation in its spectral class from F2pIb to K4e.
\item  V4728 Sgr (= HD 172481), which has the characteristics of being a Li-rich metal deficient PAGB \citep{Reyniers2001}, presents a long period of 307 days (AAVSO) and low amplitude change of $\Delta$V$\sim$0.5 mag
\item Finally R Sct (= HD 173819) is the third RV Tauri star of the sample and the brightest of its class. It shows a period  estimated between 140.2 and 146.5 days as well as a large amplitude variation $\Delta$V up to $\sim$4 mag (AAVSO, GCVS).
\end{itemize}

\section{Observations and data reduction}
\begin{table}
	\caption{\label{masks}Summary of LSD line mask properties including the effective temperature, the logarithmic surface gravity and the number of spectral lines used.}
	\begin{center}
	\begin{tabular}{l c c}
	\hline
$T_{\rm eff}$ & $\log g$ &  $\#$ lines\\
(K) & (cgs) & \\
\hline
4500 & 0.0         &             15147\\
5000 & 0.0          &            12103\\
5500 & 0.5          &            10114\\
6500 & 1.0 & 5441 \\
6500& 0.5         &             5995\\
7250 & 1.5         &             3703\\
\hline
\end{tabular}
\end{center}
\end{table}

\begin{figure*}
{\includegraphics[height=9cm]{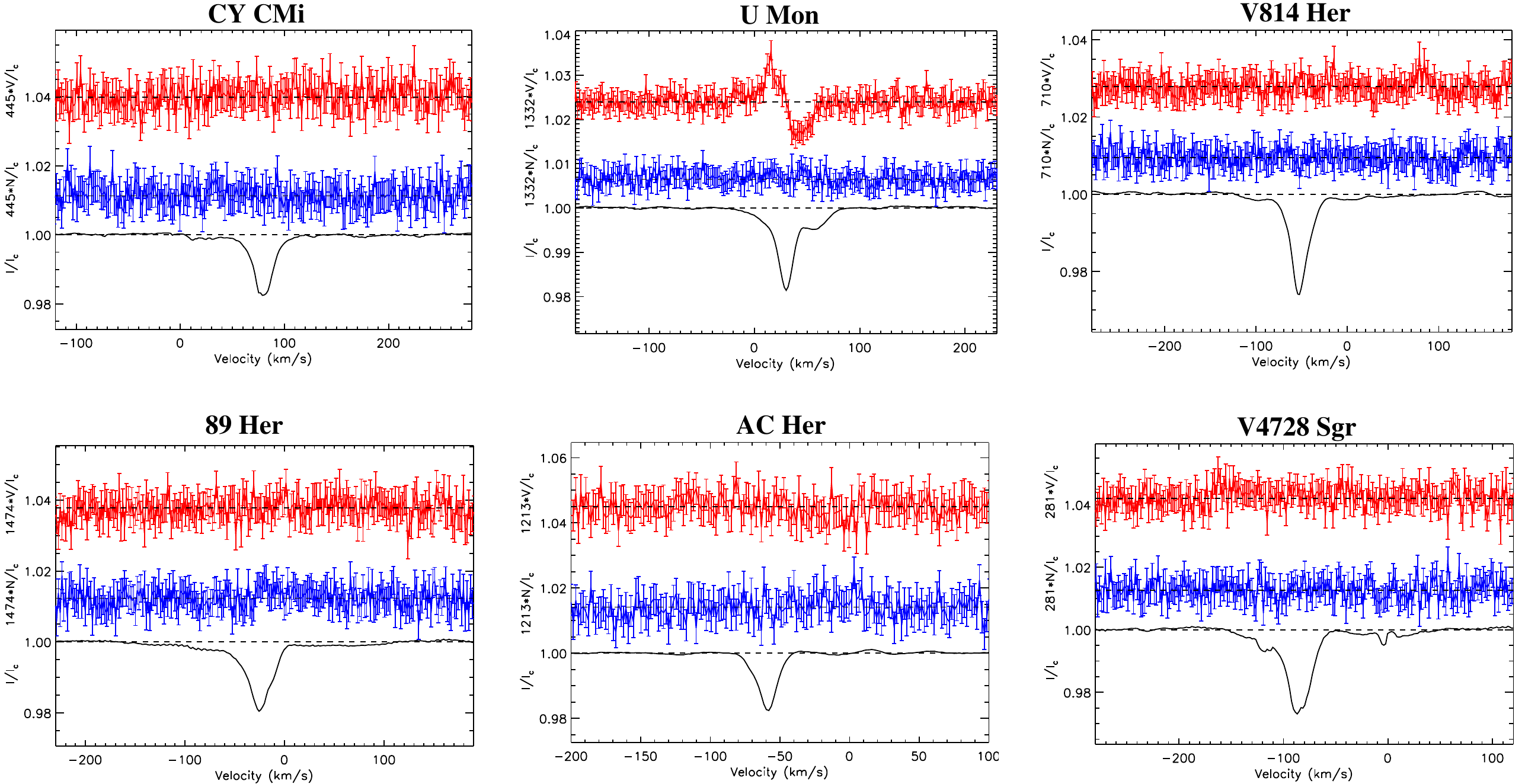}} 
\caption{\label{spectra} LSD profiles obtained with ESPaDOnS for the PAGB stars CY CMi, U Mon, V814 Her, 89 Her, AC Her and V4728 Sgr. For each object, the upper plot (in red) shows the mean Stokes $V$ profile, the middle plot (in blue) shows the null ($N$) profile and the bottom plot (in black) indicates the mean Stokes $I$ profile. The amplification factors for $N$ and $V$ differ for each star. A clear magnetic signature is seen in the PAGB U Mon.}
\end{figure*}
\begin{figure*}
{\includegraphics[height=4.1cm]{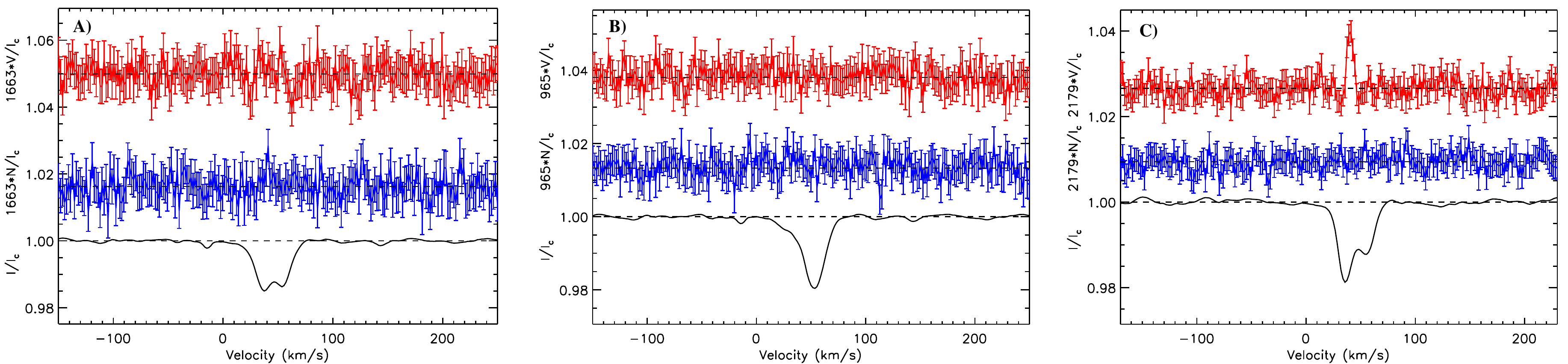}} 
\caption{\label{Rsct_spec} LSD profiles obtained with the high resolution spectropolarimeter Narval of R Sct on 05-Sept-2007  (panel A), on 03-Jun-2008 (panel B) and the average profiles obtained between the 21 and 23-July-2014 (panel C). Similarly to Fig.\ref{spectra} the data displayed are the Stokes $V$, null and Stokes $I$ profiles. The analysis of the latest set of data reveals not only a clear magnetic field detection, but also a temporal (phase related) variation of the Stokes $I$ profile linked to the shock processes occurring in R Sct (see text).}
\end{figure*}

\begin{table*}
\begin{center}
\caption[]{\label{results} Results of the LSD analysis with the magnetic field strength and its associated error, the parameters (temperature and gravity) for the LSD masks used and the corresponding detection flag (see text). We emphasize in bold the characteristics of the two objects for which a definite detection has been found.}
\hspace{-0.5cm}
\begin{tabular}{|l|c|c|c|c|c|c|c|}
\hline
 Parameters &  CY CMi   & {\bf U Mon} & V814 Her & 89 Her & AC Her & V4728 Sgr & {\bf R Sct} \\
 \hline 
B$_{\ell}$ $\pm$ $\sigma_{B}$ (G) &  -2.8$\pm$6.0  &{\bf 10.2$\pm$1.7} & -1.0$\pm$1.8 & -1.5$\pm$2.2 &  3.1$\pm$2.2 &  17.8$\pm$8.1 & {\bf 0.6$\pm$0.6} \\
LSD mask (K,log g) & 6500,0.5 & {\bf 5000,0.0} & 6500,0.5 & 6500,1.0 & 5500,0.5 & 7250,1.5 & {\bf 4500,0.0}\\
LSD flag  & ND &{\bf DD} &ND & ND& ND& ND & {\bf DD}\\
\hline
 \end{tabular}
\begin{minipage}{10.8cm}
\end{minipage}
\end{center}
\end{table*}

Spectropolarimetric observations were conducted in two runs in 2009 and in 2014 (runs ID  09AO01 and 14AC08) with the \'echelle spectropolarimeter ESPaDOnS \citep{Donati2006} located at the 3.6m Canada-France-Hawaii Telescope (CFHT, Hawaii) and in one run with Narval (TBL, France), twin of ESPaDOnS, in 2014. In both instruments, the spectral range covers a large part of the visible domain i.e. from 370 nm to 1050 nm over 40 spectral orders with a resolving power R $\sim$ 65~000. The unpolarised spectrum (Stokes $I$) and the circular polarised spectrum (Stokes $V$) were obtained in Queued Service Observing (QSO) mode following the standard procedure in which each spectropolarimetric sequence consisted of four individual sub-exposures taken in different orientations of the half-wave Fresnel rhombs. Stokes $I$ and Stokes $V$ spectra were then derived for each set of sub-exposures. The observations were mostly conducted under sub-arcsecond seeing conditions. The complete data reduction was performed with the CFHT's reduction pipeline Upena which is based on the software Libre-ESpRIT \citep{Donati1997}. This fully automatic reduction process performs classical operations on spectropolarimetric data: bias subtraction, flat-fielding, removal of bad pixels, wavelength calibration, spectrum extraction, and extraction of the polarimetric information. We also looked for any previous  spectropolarimetric observations of the stars in our sample in ``PolarBase\footnote{http://tblegacy.bagn.obs-mip.fr/}'' \citep{Petit2014}. This database contains a large number of reduced and calibrated spectropolarimetric data obtained with both  ESPaDOnS and Narval. Thus, we retrieved Stokes $Q$,$U$ and $V$ data for R Sct, which analysis will be combined to our more recent dataset (see \S4.2).\\

 Because most of PAGB stars are known to exhibit pulsations with a wide range of amplitudes, periods and cycle regularity, one of the main constraints of our program was the photometric variability of our sample targets (see $\S$2). We programmed the observations of the stars around maximum light (e.g. at their brightest phase), when their photosphere is the hottest. The time constraint for each target was derived based on literature data and light curves from the AAVSO and the GCVS.  \\

Table \ref{Obs} presents the log of our Stokes $V$ observations with the target ID, the date of observation, the number of collected $V$ sequences, the total exposure time (in sec) for a single Stokes $V$ sequence and the peak signal to noise ratio (S/N) of a single Stokes $V$ sequence per 2.6{\,${\rm km.s}^{-1}$\,} spectral bin, around 680~nm. As mentioned by \citet{Leone2014}, our ability to detect any magnetic field is strongly linked to the high S/N achieved. In our case, the mean S/N per spectral bin per Stokes $V$ sequence is typically greater than $\sim$500.\\

\section{Diagnosis of the magnetic field}

We applied the Least-Squares Deconvolution (LSD) procedure \citep{Donati1997} as implemented in the iLSD code of \citet{Kochukhov2010}. We employed line masks constructed from Vienna Atomic Line Database (VALD3, \citealt{Kupka1999}) {\tt extract stellar} requests from 3690 to 10,000~\AA~ for temperatures and gravities corresponding to our targets, with a line depth threshold of 0.2 and 2~km/s micro-turbulence. Masks were iteratively cleaned by direct comparison of the LSD model spectrum with the observed spectrum, in particular removing broad lines (H lines, the Ca~{\sc ii} H and K lines) as well as strong interstellar lines, lines in obvious emission, and lines blended with strong telluric features. Ultimately, this resulted in most lines redward of 686~nm being removed from the mask. The number of lines in the final masks are summarized in Table~\ref{masks}.\footnote{We also tested line masks with various other parameters, including a line depth threshold of 0.1 rather than 0.2, and no cleaning. Ultimately we found that this led to no significant improvement in the quality of the magnetic diagnosis.}

The LSD profiles, extracted using weights of $\bar \lambda=500$~nm, $\bar z=1.2$ and $\bar d=0.1$ (corresponding to the normalisation wavelength, Land\'e factor and line depth, respectively) are illustrated in Fig.~\ref{spectra}. Each panel shows the mean Stokes $I$ (bottom), Stokes $V$ (top) and diagnostic null (middle) profiles for one star.
We analyse the LSD profiles in two ways. First, we compute the formal $\chi^2$ detection probability of significant signal in both Stokes $V$ and the null within the bounds of the line profile \citep{Donati1997}. A definite detection (DD) is achieved for false alarm probability (fap) below 10$^{-5}$, a marginal detection (MD) for 10$^{-5}$ $<$ fap $<$ 10$^{-3}$, and a non detection (ND) for fap above 10$^{-3}$. Secondly, we use the first-order moment method \citep{Rees1979} adapted to LSD to derive the longitudinal magnetic field $B_{\ell}$ according to the equation:
\begin{equation}
\hspace{1.5cm}
B_\ell= -2.14 \times 10^{12} \frac{\int\nu V(\nu)d\nu}{{\bar \lambda} {\bar z}c\int[I_{c}-I(\nu)]d\nu} 
\end{equation}

\noindent with $\lambda$ in \AA~being the normalisation wavelength of the LSD profile, $c$ the speed of light  (in the same units as $v$) and $z$ the normalisation Land\'e factor \citep{Wade2000}. \\
The results of these analyses are reported in Table~\ref{results}. Two significant detections of signal are obtained from the V and N profiles of our targets: a definite detection in the Stokes $V$ profile of U Mon and R Sct. All other $V$ and $N$ profiles correspond to non-detections.\\ 

\subsection{U Mon}
The magnetic field in U Mon is significantly detected, and corresponds to a longitudinal magnetic field of $10.2\pm 1.7$~G (versus $1.4\pm 1.7$~G measured from the $N$ profile). As can be seen in Fig.~\ref{spectra}, the mean Stokes $I$ line profile is complex: it presents a double component structure consisting of a sharp, narrow and blueshifted component of width $\sim 15$~km/s centred at about $+30$~km/s, and a weaker, broad and redshifted component spanning the approximate range $-20-80$~km/s. This structure is typical of a dynamic atmosphere impacted by shock wave circulation, as has already been reported for U Mon by \citet{Baird1984} and \citet{Bopp1984}. \citet{Pollard1997} claimed that the large velocity spacing of the double components observed in U Mon (appearing larger than in other RV Tauri stars) could be due to the presence of a shock of greater intensity. The Stokes $V$ signature is large and spans the entire width of the double line Stokes $I$ profile; i.e. it is not restricted to the stronger blueshifted narrow profile as it is in the Mira star $\chi$ Cyg \citep{Lebre2014}.

\begin{figure*}
{\includegraphics[height=10cm]{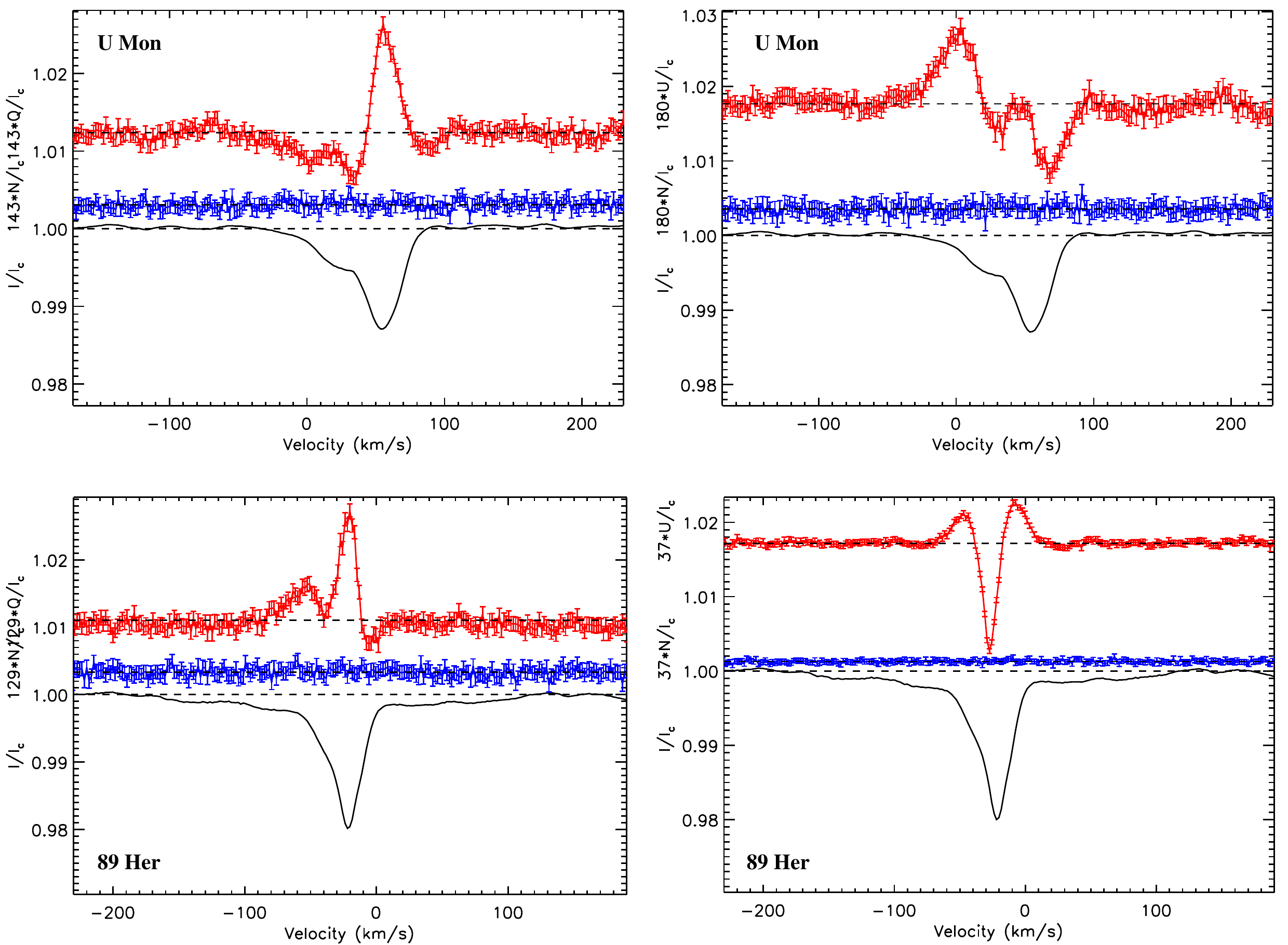}} 
\caption{\label{linear} Top row: Stokes $Q$ (left) and $U$ (right) LSD profiles of U Mon obtained with ESPaDOnS. The observations performed on 8th February 2006 were extracted from the CFHT Archive. A notable feature is the change of the double line structure of the Stokes $I$ profile when compared to our 2014 results. Indeed the redshifted and blueshifted components have an opposite behaviour likely linked to the shock propagation pattern in U Mon. The Q and U pattern can also be associated to an asymmetric morphology near the photosphere \citep{Fabas2011,Lebre2014}. Bottom row: The Stokes $Q$ (left) and $U$ (right) LSD profiles of 89 Her indicate a strong linear polarisation in this star, but the latter does not present any Stokes $V$ signature as can be seen in Fig.\ref{spectra}. This pattern supports our claim that the low level of cross-talk is not affecting our measurements.}
\end{figure*}

Although PolarBase does not contain any further observations of U Mon, a search of the CFHT Archive at the Canadian Astronomy Data Centre\footnote{\scalebox{0.85}{http://www2.cadc-ccda.hia-iha.nrc-cnrc.gc.ca/cadcbin/cfht/wdbi.cgi/cfht/quick/form}} reveals additional observations of U Mon acquired with ESPaDOnS in 2006. These were obtained in linear polarisation (Stokes $Q$ and $U$) on 8/9 February 2006. The Stokes $I$ profile in Feb 2006 (see Fig. \ref{linear}-top) also displays a double component structure but the redshifted component is much more developed than the blueshifted one.
Considering the AAVSO light curve of U Mon, we noticed that the February 2006 observations were collected while the star was reaching a deep minimum light, while our April 2014 observation took place (as planned) when the star was at maximum light (occurring after a shallow minimum light). In the former case (February 2006) and according to the shocks scenario a new shock has just emerged out from the photosphere and has started to propagate outward throughout the stellar atmosphere driving the lower atmosphere's material upward (resulting in the emergence of a shallower and blueshifted component) while a large fraction of the atmosphere is relaxing (resulting in a well developed redshifted component). In the latter case (April 2014) an intense shock has emerged from the photosphere (around the shallow minimum light), while the atmosphere was not fully relaxed after the propagation of the previous shock. This interaction is responsible for a very developed blueshifted component, while the redshifted one is receding. The extension of all observed $V$, $Q$ and $U$ profiles across the width of the entire line profile supports this interpretation, rather than e.g. a profile resulting from two distinct stellar components (i.e. an SB2 system). \\
The linear polarisation measured from the 2006 spectra using LSD is rather strong - about $7\times$ the peak-to-peak amplitude of circular polarisation we measured in 2014. However, at present the ESPaDOnS crosstalk is routinely measured, and has been found to be consistently below 0.5\% (e.g. \citealt{Silvester2012}; Manset, priv. comm.). Therefore, even if the level of linear polarisation present in 2014 was as large as that measured in 2006, crosstalk of Stokes $Q$/$U$ into Stokes $V$ cannot explain any significant fraction of the Stokes $V$ signal we measured in 2014. This conclusion is also supported by the lack of detection of Stokes $V$ signatures in other stars of our sample that have also historically exhibited strong linear polarisation (e.g. 89 Her, Fig.~\ref{linear}).\\

\subsection{R Sct}

Two Stokes $V$ sequences of R Sct are available from PolarBase. They were collected on the 5th of September 2007 (around a maximum light following a deep minimum light, considering the AAVSO light curve) and on the 3rd of June 2008 (around a deep minimum light). These observations have been obtained with a peak S/N of $\sim$1200 at $\sim$720 nm.\\  
The subsequent LSD analysis performed with a mask computed for T{\tiny{eff}} = 4500 K and log g = 0 (Fig.\ref{Rsct_spec}) led to a weak (and not formally significant) Stokes $V$ signal in the first observation (05 sept 2007, panel A), with a measured field $B_{\ell}$=4.1$\pm$2.2 G. However, there was no Stokes $V$ signal in the second observation (panel B): the associated $\chi^2$ statistic yields a non detection and a measured magnetic field $B_{\ell}$=3.1$\pm$3.1 G. \\
A more recent set of six Stokes $V$ observations of R Sct were obtained with Narval (TBL) during three consecutive nights, from the 21st to 23rd of July 2014, while the star was on a maximum light plateau (see details in Table~\ref{Obs}). The LSD Stokes $I$ and $V$ profiles, resulting from the combination of those 6 sequences, are displayed in Fig.\ref{Rsct_spec} (panel C). A clear Stokes $V$ signal is detected and the LSD analysis reports a definite detection. The measured longitudinal component of the magnetic field $B_{\ell} = 0.6\pm 0.6$~G.\\

The Stokes $I$ profiles of R Sct presented in panels A and C of a Fig.~\ref{Rsct_spec} show the double component structure typical of the presence of a shock wave propagating through the lower atmosphere. Such time variable line profiles have already been reported and monitored with the pulsation phase of R Sct  \citep{Lebre1991}. However, and differently from the case of U Mon (see \S4.1 and Fig\ref{spectra}), the Stokes $V$ profile we detect for R Sct is clearly associated with only the blueshifted component of the Stokes $I$ profile, i.e. with the material being lifted up in the wake of the shock wave. This trend was also reported in the case of the Mira star $\chi$ Cyg \citep{Lebre2014} suggesting a compressive action of the shock on a weak surface magnetic field. 

\subsection{Limits on the significant measurements}

Apart from the detection of a Stokes $V$ signal in U Mon and R Sct, no other significant detections were obtained in this study. In our sample, our observations of 3 additional stars (V814 Her, 89 Her and AC Her) yielded longitudinal field error bars comparable to those of U Mon, while 2 other stars (CY CMi and V4728 Sgr) yielded somewhat larger error bars. In the former cases, our observations would have detected a magnetic field comparable to that of U Mon. In the latter cases, we would have been insensitive to such a field. We can therefore state that magnetic fields with mean longitudinal components larger than $\sim$10~G are not generally present in the photospheres of PAGB stars. However we can not fully exclude intermittent behaviour of the magnetic field, as reported by \citet{Konstantinova2013} for AGB stars and thus spectropolarimetric monitoring of these targets is foreseen.

\section{Discussion}

\subsection{Are RV Tauri stars the ``magnetic PAGB population'' ?}

Our small but diverse sample represents a first testbed to investigate the likelihood for a PAGB star to show a detectable surface magnetic field. We unveiled such magnetic field in the PAGB stars U Mon and R Sct both of which are RV Tauri stars. Without any other lead, we can therefore wonder if this class is more likely to display surface fields.\\
The other studied RV Tauri star AC Her, which shows a non detection, shares some similarities with U Mon: they are both long period binaries: P= 1194$\pm$6 days and {\it a $\sin$i}= 1.39 AU for AC Her \citep{Winckel1998}; P $\sim$2597 days and {\it a $\sin$i}= 3.6 AU for U Mon  \citep{Pollard1995,Ruyter2006}; they also both have low mass discs: $<10^{-3} M\sun$ \citep{Bujarrabal2013}. However they differ in photometric and spectroscopic classes, U Mon being classified as RVb/RVA and AC Her as RVa/RVB \footnote{RV Tauri stars are classified based on photometric and spectroscopic factors. In the first case, photometric classification focused on whether the stars show or not long periodic variation in their maximum light and are named RVb and RVa respectively. In the second case, the spectroscopic classification, namely RVA, RVB and RVC, relies on the temperature and chemical composition of the stars.}; in their metallicity ($\left[Fe/H\right]$=-0.8 for U Mon against $\left[Fe/H\right]$=-1.5 for AC Her) and in terms of effective temperatures (see Table \ref{Sample}). Finally RV Tauri stars are known to show abundance peculiarities (e.g. no high C and s-process abundances) which are linked to the separation of dust from the gas in a circumstellar disc \citep{Waters1992}. This depletion pattern (of refractory elements) is more pronounced in AC Her (C-rich) than it is in U Mon (O-rich) \citep{Giridhar2000}. Ultimately, we found no obvious parameter(s), linked to the basic characteristics of U Mon and AC Her, which would explain the difference in their magnetic fields. \\


Contrary to U Mon, R Sct has not been found to be part of a binary system \citep{Ruyter2005}. The latter authors also pointed out the much lower dust excess in R Sct compared to U Mon with drastically different spectral energy distributions. However, \citet{Bujarrabal2013} derived much higher total nebular and disc masses for R Sct with values of $5\times10^{-2} M\sun$ and $\sim 6.9\times10^{-3} M\sun$ respectively. Other comparative points are the metallicity (higher in R Sct), the effective temperature (lower in R Sct) and the photometric  and spectral classes which are respectively different and similar in both PAGBs. In terms of chemistry, \citet{Giridhar2000} noted that the dust-gas separation in R Sct (O-rich) is limited to certain elements (with the highest condensation temperatures) and the pattern is not as strong as in the RVB AC Her for example. \\

\indent In conclusion, except for the fact that the two detections of surface magnetic fields were made in O-rich RVA-type RV Tauri stars (i.e. with generally lower effective temperatures and less affected by the dust-gas separation following \citealt{Giridhar2000}), there is no obvious parameter that can immediately be linked to the presence of a surface magnetic field. However we must emphasise on the small size of the sample which hampers any solid statistical analysis and also on the fact that we can not exclude any intermittent behaviour of such surface magnetic fields, as it has been reported for AGB stars \citep{Konstantinova2013}.  \\

\subsection{Influence of the weak field on the dynamics}
As stated in $\S$1, one reason to look for magnetic fields in evolved objects is connected to the ability of these fields to help shape the envelopes surrounding the stars (alongside a binary companion) by launching and collimating the flows. 
The asymmetries which appear as early as the AGB phase (e.g. \citealt{Mauron2006}) are well seen in the pre-PN and Planetary Nebula phases \citep{Sahai2011,Balick2007}. Magneto-Hydrodynamical (MHD) models by e.g. \citet{Garcia1999} and \citet{Gardiner2001} have shown the capability of weak fields to shape the nebulae of objects in their post AGB phase. MHD modelling of AGB stars by \citet{Pascoli2010} for example predicted that surface magnetic fields as low as 10--100 G would be sufficiently strong to eject the wind on the path towards the planetary nebula phase (in the classical $AGB \rightarrow PAGB \rightarrow PN$ sequence).\\
Our magnetic field measurements in the photosphere of U Mon and R Sct support the use of weak fields in MHD models.\\ 
In addition, \citet{Ruyter2005} suggested that U Mon hosts a ``non-spherically symmetric circumstellar dust distribution'' based on the comparison of its large IR excess and the lack of line-of-sight reddening. The strong Stokes $Q$ and $U$ profiles presented in Fig.~\ref{linear} seem to support this claim. The presence of a bipolar envelope in R Sct has been discussed by \citet{Buj1990,Buja2001} through carbon monoxide observations.\\
The exact scenario describing how the weak magnetic field found at the surface of U Mon and R Sct is implicated in the stars dynamics is still unclear. But the coexistence in the RV Tauri stars of a disc, a (long period) binary system [except for R Sct] and a surface magnetic field are promising elements in the study of any ``sculpting mechanism''. In Both cases, further MHD analysis is then necessary to clarify the combined effect of all these components in the shaping process.\\
We also emphasize the need for further study of the binary system as the nature of the companion of U Mon is still unknown (as it is the case for several PAGB stars).\\



\begin{figure}
\hspace{-0.4cm}
{\includegraphics[height=6cm]{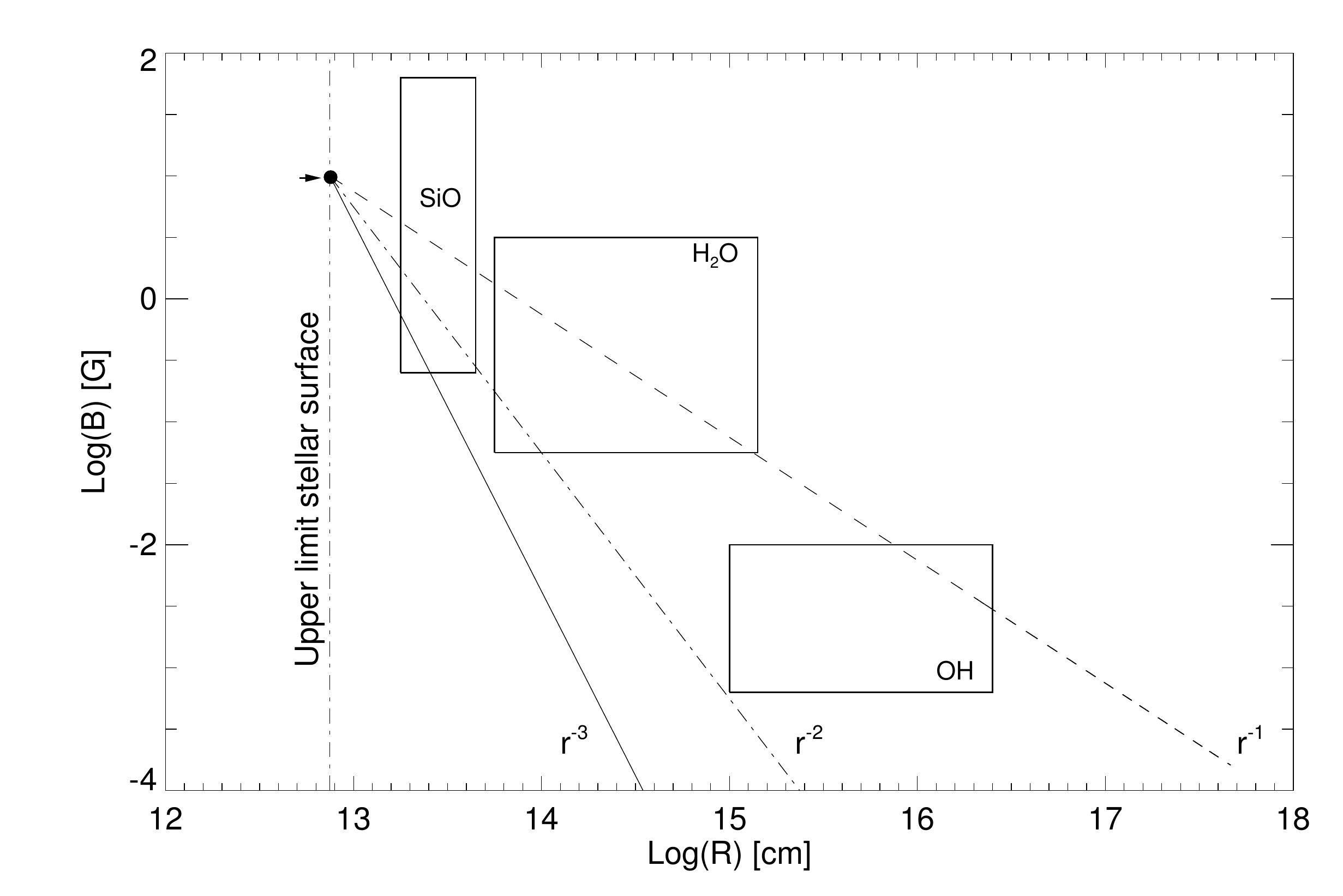}} 
\caption{\label{maser} Variation of the magnetic field as a function of the distance from the stellar surface of U Mon. We present the variation of the expected field strength for each magnetic configuration: B $\propto$ r$^{-1}$, B $\propto$ r$^{-2}$, B $\propto$ r$^{-3}$ for a toroidal, poloidal/solar or dipole structure respectively. The arrow and black circle indicate our measurement. We superimposed the approximate location of the SiO, H$_{2}$O and OH masers based on the data from \citet{Vlemmings2011}. Such a graph will therefore have to be compared with polarimetric Stokes $V$ data from different maser species to check if they point towards a local or global nature of magnetic fields and determine the characteristics of the medium in which the magnetic field evolves.}
\end{figure}

\subsection{Prediction of maser strengths in the envelope of U Mon}

In $\S$1 we mentioned the method which consists in using measurements of different maser species to extrapolate the surface magnetic field by assuming a particular field configuration. Our new data will allow us to perform the reverse analysis: for each configuration we will be able to predict the strength of the field given by the SiO, H$_{2}$O and OH masers {\it assuming} first an homogeneous environment and then that the masers describe a global magnetic field, i.e. they are not the source of localized magnetic fields unrelated to the global field issued from the central star. Fig.\ref{maser} includes previous measurements of SiO masers \citep{Kemball2009,Herpin2006}, H$_{2}$O masers \citep{Vle2002,Vlemmings2005} and OH masers \citep{Rudnitski2010} giving us an overview of the maser and  magnetic field strength distributions in evolved stars. Thus, \citet{Vlemmings2011} summarised the results of the aforementioned studies related to the characteristics of masers in AGB stars (the behaviour is similar in PAGB stars) and noted some correlations through the different investigations in terms of strength and location of each masers. Hence, SiO masers show intensities of $\sim$3.5 G and are located between 2--4 AU from the stellar surface; H$_{2}$O masers present a general magnetic strength of $\sim$0.3 G and are located in a range 5--50 AU and finally OH masers have intensities of $\sim$3$\times$10$^{-3}$ G and are present between 100--10000 AU from the surface.\\ 
We decided to focus on U Mon as it shows the clearest and strongest magnetic signature. We therefore followed the method described by \citet{Ferreira2013} and derived an upper limit on the stellar radius of $\sim$0.5 AU based on the luminosity and effective temperature given by \citet{Ruyter2005}. The magnetic fields measured at the surface of U Mon allows us to extrapolate and predict that: for OH masers (probing the external part of the circumstellar envelope) the maximum magnetic field expected assuming a B $\propto$ r$^{-1}$ law would be of the order of 10 mG; for H$_{2}$O masers following a B $\propto$ r$^{-1}$ one would expect a maximum field strength of $\sim$1.6 G and $\sim$0.2 G assuming a B $\propto$ r$^{-2}$ law. Finally SiO masers, which can fit all three configurations, would display maximum values of about 4 G, 2 G and 0.4 G for B $\propto$ r$^{-1}$, r$^{-2}$ and r$^{-3}$ respectively. In the case of the SiO masers all the laws may be compatible within the associated error bar making OH masers the best discriminant. Nevertheless, the detection and analysis of OH masers remains a complex operation and requires high sensitivity/resolution for both the line detection and subsequent (spectro)polarimetric study. The resulting graph, shown in Fig.~\ref{maser}, is therefore a useful tool as we have now an indication of the expected magnetic field for each maser species. \\ 
 It is worth mentioning that the Stokes $V$ line profiles measured in the photospheric spectrum of U Mon corroborate the presence of a Zeeman effect. This has to be taken into account when analysing SiO maser polarization for example as non-Zeeman circular polarization signals (leading to a different interpretation of the masers polarization) also exist as shown by \citet{Houde2014}.\\ 
\indent In the case of U Mon, different studies, compiled by \citet{Benson1990}, indicated that no OH nor H$_{2}$O masers were detected. \citet{Planesas1991} concluded that the lack of detected OH masers in RV Tauri stars they observed could be linked to a deficiency of molecules in their envelopes. But the authors also pointed out that strong OH lines should be seen due to the strong Far--IR emission of U Mon and the non detection would be due to the inherent weak maser emission in RV Tauri stars compared to Mira-type stars. We also have to take into account the temporal variability of masers which can appear and/or disappear from one epoch to the other as it was recently shown by \citet{Vlemmings2014}. Therefore a new observing campaign combined with the use of sensitive interferometric facilities such as ALMA (when its line polarimetric capability will be available) would help to test our predictions by allowing the detection and measurement of the masers at various depths in the envelope.\\



\section{Summary}

The precise estimation of the impact of magnetic fields on the dynamics (and chemistry) of PAGBs and PNe is mostly hampered by  the lack of surface magnetic fields measurements.\\
In order to tackle this problem, we therefore opted to take advantage of the high resolution spectropolarimeters ESPaDOnS and Narval (R $\sim$ 65 000) and to observe a well-chosen and diverse sample of PAGB stars. Indeed, the objects are bright with well defined, numerous and sharp absorption profiles and little or no nebular contamination. They are also from different PAGB classes and finally they are either single or in a binary system. The data analysis was performed using the Least-Squares Deconvolution (LSD) technique which renders the Stokes $V$, $I$ and null profiles as well as the longitudinal surface magnetic field $B_{\ell}$. \\

We therefore report the \textit{\textbf{first clear discovery of weak magnetic fields at the stellar surface of two PAGB stars}} namely U Mon and R Sct. The LSD analysis gives in both cases a definite detection flag and an associated longitudinal magnetic field of $10.2\pm 1.7$~G for U Mon and $0.6\pm 0.6$~G for R Sct. The Zeeman signature is very clear in the Stokes $V$ profiles of both sources and is not affected by any significant crosstalk effect.\\

The fact that the two ``magnetic stars'' U Mon and R Sct belong to the RV Tauri class combined with the presence in our sample of an undetected RV Tauri star (AC Her), led us to look for any commonalities or striking differences in these stars which could be associated with the presence (or absence) of magnetic fields in this class. Among the different tested criteria none seemed to be a clear discriminant. However other parameters such as stellar rotation still have to be studied.\\
The observed characteristics of the line profiles and Zeeman signature suggest that the magnetic field is associated with the dynamical state of the atmosphere, and will vary with the pulsational phase of the star. Spectropolarimetric monitoring is now foreseen in order to explore the relation of the magnetic field and the atmospheric dynamics. Another possible interpretation of the combined ‘broad-narrow’ profile of U Mon in particular, is an SB2 system. We consider this possibility to be unlikely, given the argument stated above. Testing the SB2 scenario will require more data.\\
Finally, another output from our study on U Mon in particular, is the ability to predict the strength of different masers, at various distances in the  envelopes for different field configurations. We were therefore able to estimate intensities ranging from $\sim$4 G to $\sim$10 mG for SiO, H$_{2}$O and OH masers. Due to the weak molecular emission from RV Tauri stars, highly sensitive instrumentation will be needed to test our prediction.\\

The detections presented in this article are a first step in the investigation of magnetic fields as mechanisms playing an important role in the structure and dynamics of the envelopes of PAGBs. If data indicate an asymmetry in the dust shell or envelope of U Mon and R Sct, MHD modelling will be required to know to which extent the field could be responsible.\\ 


\section*{Acknowledgements}
The authors thank the referee for the useful comments that help improving the paper. 
LS thank Hans van Winckel for the helpful discussions on RV Tauri stars. LS is supported by the CONACYT grant CB-2011-01-0168078 and this research was also partially supported by the PIFI-2014 program from the University of Guadalajara (Mexico). GAW acknowledges a Discovery Grant support from the Natural Sciences and Engineering Research Council (NSERC) of Canada as well as the use of IDL tools developed by V. Petit and J. Grunhut. The authors acknowledge the direction and the staff of TBL for Director's Discretionary Time. This research is based on observations obtained at the Canada-France-Hawaii Telescope (CFHT) which is operated by the National Research Council of Canada, the Institut National des Sciences de l'Univers of the Centre National de la Recherche Scientique of France (INSU/CNRS), and the University of Hawaii-USA, and at the T\'elescope Bernard Lyot (TBL, at Observatoire du Pic du Midi, France) operated by the Observatoire Midi-Pyr\'en\'ees, Universit\'e de Toulouse, and INSU/CNRS.

\bibliographystyle{mn2e}

\bibliography{sabin}

\bsp

\label{lastpage}

\end{document}